\documentclass[12pt]{iopart}
\usepackage{iopams}  

\usepackage{graphicx,psfrag,epsfig}

\begin{document}

\title[Border Collision Bifurcations $\cdots$ Discontinuous
Maps]{Border Collision Bifurcations in $n$-Dimensional Piecewise
Linear Discontinuous Maps}

\author{Partha Sharathi Dutta\dag , Bitihotra Routroy\ddag , Soumitro
Banerjee\P \footnote[3]{To whom correspondence should be addressed} \
and S. S. Alam\dag}

\address{\dag\ Department of Mathematics and Centre for Theoretical Studies}

\address{\ddag\  Department of Electrical Engineering}

\address{\P\ Department of Electrical Engineering and Centre for
Theoretical Studies }

\address{Indian Institute of Technology, Kharagpur-721302, India.}

\eads{\mailto{parthas@cts.iitkgp.ernet.in},
\mailto{btroutroy@gmail.com}, \mailto{soumitro@iitkgp.ac.in} and
\mailto{alam@maths.iitkgp.ernet.in}}

\begin{abstract}

In this paper we report some important results that help in analizing
the border collision bifurcations that occur in $n$-dimensional
discontinuous maps. For this purpose, we use the piecewise linear
approximation in the neighborhood of the plane of
discontinuity. Earlier, Feigin had made a similar analysis for general
$n$-dimensional piecewise smooth continuous maps. Proceeding along
similar lines, we obtain the general conditions of existence of
period-1 and period-2 fixed points before and after a border collision
bifurcation. The application of the method is then illustrated using a
specific example of a two-dimensional discontinuous map.

\end{abstract}



\maketitle

\section{Introduction}
In recent times a significant amount of research effort has been
directed towards nonsmooth bifurcations because a large number of
physical and engineering systems have been found to exhibit this class
of bifurcation phenomena \cite{mybook,tsebook,zhusubaliyev-book}. Such
physical systems are generally given by two or more sets of
differential equations, and the system switches from one set to
another when some switching conditions are satisfied. It has been
shown that such systems give rise to piecewise smooth maps, where the
discrete-time state space is divided into two or more compartments
with different functional forms of the map, separated by borderlines
\cite{pesc97,Guohui97}. A new type of bifurcation, called {\em border
collision bifurcation} occurs when a fixed point collides with a
borderline, resulting in an abrupt change in the Jacobian matrix.
Most past studies on border collision bifurcations have concentrated
on piecewise smooth maps that are continuous across the borderlines
\cite{Nusse92,Nusse95,1d98,pre2d,cas2d,feigin-mario}.

However, it has also been found that some switching dynamical systems,
for example, the Colpitts oscillator \cite{maggio2000}, dc-dc
converters \cite{missedswitch}, thyristor controlled reactor circuits
\cite{Dobson96b}, sigma-delta modulators \cite{sigma-delta} and many
other electronic circuits \cite{discontcircuit} give rise to
discontinuous maps. Discontinuity in the map arises if, in the $n$
dimensional Poincar\'e section there exists an $(n-1)$ dimensional
manifold such that infinitesimally close points at the two sides of
the manifold map to points that are far apart.  In order to explain
the bifurcation phenomena in such systems, it is necessary to have a
bifurcation theory for discontinuous maps.

The bifurcation theory for 1D and 2D continuous piecewise smooth maps
has been developed in \cite{1d98,pre2d}. In another important line of
work, Feigin obtained some results on the existence of periodic orbits
in $n$-dimensional piecewise smooth continuous maps, which were
brought to the notice of the English speaking audience in
\cite{feigin-mario}. The theory for discontinuous maps is in the
preliminary stage of development, with some progress reported for
one-dimensional maps in \cite{parag03,kollar-disc}.

In the present paper, we follow Feigin's approach in case of discontinuous
maps, and present new results that are applicable to the general
$n$-dimensional case. In particular, we present the conditions of
existence of the period-1 and period-2 fixed points which form the
basis of the observable bifurcation phenomena in such maps.
For this purpose, we use a piecewise linear normal form where the
discontinuity is defined by two parameters.

Finally, the application of these results is illustrated
by deducing the bifurcation behavior of a two-dimensional
discontinuous map, and validating the results by simulation.

\section{The Normal Form of the Discontinuous Map}

We consider discontinuous maps of the form 
\[\bar{{\bf X}}^{(k+1)}= \left\{ \begin{array}{ccc} 
\Pi^- (\bar{{\bf X}}^k) & \mbox{for} & \bar{{\bf X}} \in P^-\\
\Pi^+ (\bar{{\bf X}}^k) & \mbox{for} & \bar{{\bf X}} \in P^+\\
			   \end{array} \right.
\]
where the discrete-time phase space is divided into two regions $P^-$
and $P^+$, and $\Pi^-$ and $\Pi^+$ represent the functional forms of the
map in the two regions respectively.

Since our points of interest are the occurrence of bifurcations with the 
traversal of a fixed point across the border, we can study the same by
applying local linear approximation in neighborhood of the 
discontinuity:
\begin{equation}
{\bf X}^{(k+1)} = F({\bf X}^{(k)}) = \left\{\begin{array}{cc} {\bf
A}_1 {\bf X}^{(k)} + {\bf C} \mu_1 , & \mbox{if}\; x_1^{(k)} < 0\\
{\bf A}_2{\bf X}^{(k)} + {\bf C} \mu_2, & \mbox{if}\; x_1^{(k)} > 0
\end{array} \right.
\label{normal-nd}
\end{equation}       
where $${\bf A}_1=\left.\frac{\partial\Pi^{-}}{\partial{\bf
X}}\right|_{{\bf X}=0},\;\;\; {\bf
A}_2=\left.\frac{\partial\Pi^{+}}{\partial{\bf X}}\right|_{{\bf
X}=0}$$ and
$${\bf C}=\left.\frac{\partial\Pi^{-}}{\partial\mu_1}\right|_{\mu_1=0} =
\left.\frac{\partial\Pi^{+}}{\partial\mu_2}\right|_{\mu_2=0}.$$ 
Here ${\bf X}^{(k+1)}$ is $(k+1)$th iteration of
$n \times 1$ matrix ${\bf X}=(x_1,x_2, \dots ,x_n)^{T}$, ${\bf A}_1$
and ${\bf A}_2$ are $n \times n$ matrices, ${\bf C}=(c_1,c_2, \dots
,c_n)^{T}$ is $n \times 1$ matrix and $\mu_1,~ \mu_2$ are parameters.

We can apply a suitable coordinate transformation to obtain the
matrices in the normal form, where the matrices ${\bf
A}_1=[a_{ij}^{(1)}]$ and ${\bf A}_2=[a_{ij}^{(2)}]$ are such that
$a_{ij}^{(1)} = a_{ij}^{(2)}$ if $j \neq 1$, i.e., except the first
columns of the matrices all other columns are same \footnote{The
derivation of the normal form in one-dimension can be found in
\cite{1d98}, and that in two-dimension can be found in
\cite{guohuithesis, pre2d}}. In the earlier work on piecewise smooth
continuous maps \cite{feigin-mario,pre2d} this condition was required to ensure
continuity of the map. In the present work, though we are considering
discontinuous maps, we retain this property of the matrices ${\bf
A}_1$ and ${\bf A}_2$ and move the discontinuity to the parameters
$\mu_1$ and $\mu_2$ without loss of generality. In \cite{parag03} the
length of the discontinuity was considered to be constant. In this
paper, in order to allow a variable length of the discontinuity, we
have used two parameters $\mu_1$ and $\mu_2$.  We have assumed ${\bf C}$ to
be same on both sides, which physically means that the two parameters
have proportional effect on the two sides of the border.

\subsection{Existence of period-1 fixed points}

Let us consider two fixed points ${\bf X}^{*}_{L} = (x_{1L}^{*},x_{2L}^{*},
\dots, x_{nL}^{*})^{T}$ and ${\bf X}^{*}_R= (x_{1R}^{*},x_{2R}^{*}, \dots,
x_{nR}^{*})^{T}$ of the sub-mappings $\Pi^{-}$ and $\Pi^{+}$
respectively. That is,
\begin{equation}
{\bf X}^{*}_{L} = {\bf A}_1 {\bf X}^{*}_{L} + {\bf C} \mu_1 ~~~ \mbox{for} ~~~
x_{1L}^{*} < 0,
\end{equation}
\begin{equation}
{\bf X}^{*}_R = {\bf A}_2 {\bf X}^{*}_R + {\bf C} \mu_2 ~~~ \mbox{for}
~~~ x_{1R}^{*} > 0.
\end{equation} 
where $x_{1L}^{*}$ and $x_{1R}^{*}$ are the first components of ${\bf
X}^{*}_L$ and ${\bf X}^{*}_R$ respectively.

We define $p^{*}{(\lambda)} = |{\lambda}{\bf I}-{\bf A}_1|$ and
$p^{**}{(\lambda)} = |{\lambda}{\bf I}-{\bf A}_2|$, where ${\bf I}$ is
the $n \times n$ identity matrix \footnote{Note that Feigin
\cite{feigin-mario} defined these in the opposite way, as
$p^{*}{(\lambda)} = |{\bf A}_1-{\lambda}{\bf I}|$ and
$p^{**}{(\lambda)} = |{\bf A}_2-{\lambda}{\bf I}|$.}.
                 
Assuming $({\bf I}-{\bf A}_1)$ and $({\bf I}-{\bf A}_2)$ to be
invertible, we find
\begin{center}
${\bf X}^{*}_L = ({\bf I}-{\bf A}_1)^{-1} {\bf C} \mu_1 = \frac{{\rm
adj}({\bf I}-{\bf A}_1)}{p^{*}(1)} {\bf C} \mu_1$,
\end{center}
\begin{center}
${\bf X}^{*}_R = ({\bf I}-{\bf A}_2)^{-1} {\bf C} \mu_2 = \frac{{\rm
adj}({\bf I}-{\bf A}_2)}{p^{**}(1)} {\bf C} \mu_2$.
\end{center}  
Now the $k$th element of these matrices can be written as  
\begin{equation}
x_{kL}^* = \frac{b_{kL}^*}{p^{*}(1)} \mu_1, ~~~ x_{kR}^* =
\frac{b_{kR}^*}{p^{**}(1)} \mu_2\label{gen1m1m2}
\end{equation}
where $b_{kL}^* = [({\rm adj}({\bf I}-{\bf A}_1)) {\bf C}]_k$ is the $k$th
element of the column matrix $({\rm adj}({\bf I}-{\bf A}_1)){\bf C}$ and
$b_{kR}^* = [({\rm adj}({\bf I}-{\bf A}_2)) {\bf C}]_k$ is the $k$th
element of the column matrix $({\rm adj}({\bf I}-{\bf A}_2)){\bf C}$. Now,
since in the matrices ${\bf A}_1$ and ${\bf A}_2$, $a_{ij}^{(1)} =
a_{ij}^{(2)}$ if $j \neq 1$, the cofactor of first column of $({\bf
I}-{\bf A}_1)$ and $({\bf I}-{\bf A}_2)$ are same. Hence \[b_{1L}^* = \sum
\limits_{i=1}^{n} b_{1i}^{(1)} c_i = \sum \limits_{i=1}^{n}
b_{1i}^{(2)} c_i = b_{1R}^* = b_1^* \;\;\; \mbox{(say)}\]
where $ b_{1i}^{(1)}$ and $b_{1i}^{(2)}$, 
are the elements of the first rows of the matrices $({\rm
adj}({\bf I}-{\bf A}_1))$ and $({\rm adj}({\bf I}-{\bf A}_2))$
respectively and
$c_i$  are the elements of the column matrix {\bf C}.

\noindent Using the above in (\ref{gen1m1m2}) we obtain
\begin{equation}
x_{1L}^* = \frac{b_1^*}{p^{*}(1)} \mu_1, ~~~ x_{1R}^* =
\frac{b_1^*}{p^{**}(1)} \mu_2 \label{m1m2} 
\end{equation} 

We require $x_{1L}^* < 0$ for ${\bf X}_L^*$ to exist and $x_{1R}^* > 0$
for ${\bf X}_R^*$ to exist. With the help of above derivations we can
now attempt a classification of the possible behavior at the border
collision bifurcation.

Equation (\ref{m1m2}) points to two possibilities: 

\noindent {\bf Case 1:} $p^{*}(1)p^{**}(1)>0$. This 
 will be true if both $p^{*}(1)$ and $p^{**}(1)$ are positive or
if both are negative. 
There are two sub-cases.

\begin{description}
 
\item{Case 1a:} If $\mu_1 \mu_2 > 0$ then only one fixed point will exist.

\item{Case 1b:} If $\mu_1 \mu_2 < 0$ then either two fixed points will
  exist or no fixed point will exist.

\end{description}

\noindent {\bf Case 2:} $p^{*}(1)p^{**}(1)<0$. This will be true if
$p^{*}(1)>0$ and $p^{**}(1)<0$, or if $p^{*}(1)<0$ and $p^{**}(1)>0$.
Here also there are two sub-cases.

\begin{description}

\item{Case 2a:} If $\mu_1 \mu_2 > 0$ then either two fixed points will
  exist or there will not be any fixed point. 

\item{Case 2b:} If $\mu_1 \mu_2 < 0$ then only one fixed point will exist.

\end{description}

Table~\ref{table1} and Table~\ref{table2} summarize these results for
$b^*_1>0$ and $b^*_1<0$ respectively. From these tables the bifurcation
sequences can be inferred for a given set of parameters, as $\mu_1$
and $\mu_2$ are varied.


\begin{table}
\caption {\label{table1}The conditions of existence of fixed points
when $b^*_1>0$.}
\begin{indented}
\item[]\begin{tabular}{|c|c|c||c|c|c|c|} \hline
\multicolumn{3}{|c||}{Conditions on the parameters} &
\multicolumn{4}{c|}{Existence of fixed points} \\ \hline \hline
$p^*(1)p^{**}(1)$ & $p^*(1)$ & $p^{**}(1)$ & $\mu_1>0$ &
$\mu_1>0$ & $\mu_1<0$ & $\mu_1<0$ \\ 
 & & & $\mu_2<0$ & $\mu_2>0$ & $\mu_2>0$ & $\mu_2<0$\\ \hline
\hline $>0$ & $<0$ & $<0$ & ${\bf X}^{*}_L$ \& ${\bf X}^{*}_R$ & ${\bf
X}^{*}_L$ & None & ${\bf X}^{*}_R$ \\ \hline $<0$ & $>0$ & $<0$ &
${\bf X}^{*}_R$ & None & ${\bf X}^{*}_L$ & ${\bf X}^{*}_L$ \& ${\bf
X}^{*}_R$ \\ \hline $>0$ & $>0$ & $>0$ & None & ${\bf X}^{*}_R$ &
${\bf X}^{*}_L$ \& ${\bf X}^{*}_R$ & ${\bf X}^{*}_L$ \\ \hline $<0$ &
$<0$ & $>0$ & ${\bf X}^{*}_L$ & ${\bf X}^{*}_L$ \& ${\bf X}^{*}_R$ &
${\bf X}^{*}_R$ & None \\ \hline
\end{tabular}
\end{indented}
\end{table}


\begin{table}
\caption {\label{table2}The conditions of existence of fixed points
when $b^*_1<0$.}
\begin{indented}
\item[]\begin{tabular}{|c|c|c||c|c|c|c|} \hline
\multicolumn{3}{|c||}{Conditions on the parameters} &
\multicolumn{4}{c|}{Existence of fixed points} \\ \hline \hline
$p^*(1)p^{**}(1)$ & $p^*(1)$ & $p^{**}(1)$ & $\mu_1>0$ &
$\mu_1>0$ & $\mu_1<0$ & $\mu_1<0$ \\ 
 & & & $\mu_2<0$ & $\mu_2>0$ & $\mu_2>0$ & $\mu_2<0$ \\ \hline
\hline $>0$ & $<0$ & $<0$ & None & ${\bf X}^{*}_R$ & ${\bf X}^{*}_L$
\& ${\bf X}^{*}_R$ & ${\bf X}^{*}_L$ \\ \hline $<0$ & $>0$ & $<0$ &
${\bf X}^{*}_L$ & ${\bf X}^{*}_L$ \& ${\bf X}^{*}_R$ & ${\bf X}^{*}_R$
& None \\ \hline $>0$ & $>0$ & $>0$ & ${\bf X}^{*}_L$ \& ${\bf
X}^{*}_R$ & ${\bf X}^{*}_L$ & None & ${\bf X}^{*}_R$ \\ \hline $<0$ &
$<0$ & $>0$ & ${\bf X}^{*}_R$ & None & ${\bf X}^{*}_L$ & ${\bf
X}^{*}_L$ \& ${\bf X}^{*}_R$ \\ \hline
\end{tabular}
\end{indented}
\end{table}

\section{Conditions for the existence of period-2 orbit\label{secIII}}

Suppose that a new two period solution originates at a border
collision point. Let the two points of the period-2 orbit be ${\bf
M}^{*} = (m_1^{*},m_2^{*}, \dots, m_n^{*})^{T}$ and ${\bf M}^{**}=
(m_1^{**},m_2^{**}, \dots, m_n^{**})^{T}$ lying on the half planes
$L$ and $R$ respectively.

Now we derive the condition for the existence of the period-2
orbit. Using the local map (\ref{normal-nd}), we can write

\[ f({\bf M}^{*}) = {\bf M}^{**}\;\;\; \mbox{and} \;\;\; f({\bf M}^{**})
 = {\bf M}^{*}\] i.e.,
\begin{eqnarray}
{\bf M}^{*} &=& {\bf A}_2 {\bf M}^{**} + {\bf C} \mu_2, \label{m*}\\
{\bf M}^{**} &=& {\bf A}_1 {\bf M}^{*} + {\bf C} \mu_1. \label{m**}
\end{eqnarray}

Now subtracting (\ref{m*}) from (\ref{m**}) we get
\begin{equation}
\Delta {\bf M} = {\bf A}_1 {\bf M}^{*} - {\bf A}_2 {\bf M}^{**} + {\bf
C} (\mu_1 - \mu_2)\label{deltaM}
\end{equation} 
where $\Delta {\bf M} = {\bf M}^{**}-{\bf M}^{*}=
(m_1^{**}-m_1^{*},m_2^{**}- m_2^{*}, \dots, m_n^{**}- m_n^{*})^{T}=
(\delta m_1,\delta m_2, \dots, \delta m_n)^{T}$.

Now as $a_{ij}^{(1)} = a_{ij}^{(2)} = a_{ij}$ if $j \neq 1$, writing
(\ref{deltaM}) in scalar form we obtain:
\begin{equation}
\delta m_k = a_{k1}^{(1)}m_1^{*}-a_{k1}^{(2)}m_1^{**}+
c_{k}(\mu_1-\mu_2)+\sum \limits_{j=2}^{n}
a_{kj}(m_j^{*}-m_j^{**}) ,\label{deltamk}
\end{equation}
for $k=1,2, \dots,n$.

Equation (\ref{deltamk}) can be rewritten by adding and subtracting
$a_{k1}^{(1)}m_1^{**}$ as
\[
\delta m_k = (a_{k1}^{(1)}-a_{k1}^{(2)})m_1^{**}+
c_{k}(\mu_1-\mu_2)+\sum \limits_{j=1}^{n}
a_{kj}^{(1)}(m_j^{*}-m_j^{**}).
\]
Similarly by adding and subtracting $a_{k1}^{(2)}m_1^{*}$ in equation
(\ref{deltamk}) we get
\[
\delta m_k = (a_{k1}^{(1)}-a_{k1}^{(2)})m_1^{*}+
c_{k}(\mu_1-\mu_2)+\sum \limits_{j=1}^{n}
a_{kj}^{(2)}(m_j^{*}-m_j^{**}).
\]
The above two equations can be written in vector form as
\begin{equation}
\Delta {\bf M} + {\bf A}_1 \Delta {\bf M} = [{\bf A}_1-{\bf
A}_2]_{1}\: m_1^{**} + {\bf C} (\mu_1-\mu_2)\label{delMA1}
\end{equation}
\begin{equation}
\Delta {\bf M} + {\bf A}_2 \Delta {\bf M} = [{\bf A}_1-{\bf
A}_2]_{1}\: m_1^{*} + {\bf C}(\mu_1-\mu_2)\label{delMA2}
\end{equation}
where $[{\bf A}_1-{\bf A}_2]_{1}$ indicate the first column of $[{\bf
A}_1-{\bf A}_2]$.

Assuming $({\bf I}+{\bf A}_2)$ to be invertible, from (\ref{delMA2}) we obtain:
\begin{equation}
\Delta {\bf M} = \frac{{\bf B}^{*}}{p^{**}(-1)}m_1^{*} +
\frac{{\bf D}^{*}}{p^{**}(-1)}(\mu_1-\mu_2)
\end{equation}
where ${\bf B}^{*}=(-1)^{n}(\mbox{adj} ({\bf I}+{\bf A}_2))[{\bf
A}_1-{\bf A}_2]_1$ and ${\bf D}^{*}=(-1)^{n}(\mbox{adj} ({\bf I}+{\bf
A}_2)) {\bf C}$.  Similarly assuming $({\bf I}+{\bf A}_1)$ to be
invertible, from (\ref{delMA1}) we obtain
\begin{equation}
\Delta {\bf M} = \frac{{\bf B}^{**}}{p^{*}(-1)}m_1^{**} +
\frac{{\bf D}^{**}}{p^{*}(-1)}(\mu_1-\mu_2)
\end{equation}
where ${\bf B}^{**}=(-1)^{n}(\mbox{adj}({\bf I}+{\bf A}_1))[{\bf A}_1-{\bf
A}_2]_1$ and ${\bf D}^{**}=(-1)^{n}(\mbox{adj}({\bf I}+{\bf A}_1)) {\bf C}$.

Now writing (12) and (13) in scalar form

\begin{equation}
\delta m_k = \frac{b_k^{*}}{p^{**}(-1)}m_1^{*} +
\frac{d_k^{*}}{p^{**}(-1)}(\mu_1-\mu_2)
\end{equation}
\begin{equation}
\delta m_k = \frac{b_k^{**}}{p^{*}(-1)}m_1^{**} +
\frac{d_k^{**}}{p^{*}(-1)}(\mu_1-\mu_2).
\end{equation}
Since $a_{kj}^{(1)} = a_{kj}^{(2)} = a_{kj}$ for $j \neq 1$ and $k=1,2,
\dots,n$, we have $b_{1}^{**} = b_{1}^{*} = b_1$ and $d_{1}^{**} =
d_{1}^{*} = d_1$. Hence from (14) and (15)
\begin{equation}
\frac{b_1}{p^{**}(-1)}m_1^{*} + \frac{d_1}{p^{**}(-1)}\mu_0 =
\frac{b_1}{p^{*}(-1)}m_1^{**} + \frac{d_1}{p^{*}(-1)}\mu_0,
\end{equation}
where $\mu_0=(\mu_1-\mu_2)$. From (14)
\begin{equation}
m_k^{**} = m_k^{*} + \frac{b_k^{*}}{p^{**}(-1)}m_1^{*} +
\frac{d_k^{*}}{p^{**}(-1)}\mu_0,
\end{equation}
substituting (17) into (7), we obtain (in scalar form)

\[
m_k^* = \sum \limits_{j=1}^{n} a_{kj}^{(1)}m_j^{*}-
\frac{b_k^{*}}{p^{**}(-1)}m_1^{*}- \frac{d_k^{*}}{p^{**}(-1)}\mu_0 +
c_{k}\mu_1\]
or equivalently using matrices:
\begin{equation}
{\bf M}^{*} = \hat{{\bf A}}_{1} {\bf M}^{*} + {\bf C} \mu_1
-\frac{{\bf D}^{*}}{p^{**}(-1)}\mu_0
\end{equation}
where 
\begin{equation}
\hat{{\bf A}}_{1} = {\bf A}_{1} - \frac{1}{p^{**}(-1)} \left(
\begin{array}{llcl} b_1^{*} & 0 & \cdots & 0 \\ b_2^{*} & 0 & \cdots &
0 \\ \multicolumn{4}{c}\dotfill \\ b_n^{*} & 0 & \cdots & 0
\end{array}\right) \label{matrix1}.
\end{equation}

\noindent Assuming $({\bf I}-\hat{{\bf A}}_1)$ to be invertible, we
can now solve 
${\bf M}^*$. From (18)
\begin{equation}
{\bf M}^{*} = ({\bf I}-\hat{{\bf A}}_{1})^{-1}\left({\bf C} \mu_{1} -
\frac{{\bf D}^{*}}{p^{**}(-1)}\mu_0\right).\label{finalM*}
\end{equation}
Now 
\begin{equation}
|{\bf I}-\hat{{\bf A}}_{1}|=\frac{|{\bf I} p^{**}(-1)-{\bf
 A}_{1}p^{**}(-1)+{\bf L}_1|}{p^{**}(-1)}=
 \frac{d}{p^{**}(-1)},\label{detAhat}
\end{equation}
where $${\bf L}_{1} = \left( \begin{array}{llcl} b_1^{*} & 0 & \cdots & 0 \\
b_2^{*} & 0 & \cdots & 0 \\ \multicolumn{4}{c}\dotfill \\ b_n^{*} & 0
& \cdots & 0 \end{array}\right)$$ and $d$ is some constant.

\noindent Hence from (\ref{finalM*}) and (\ref{detAhat}) we obtain 
\begin{eqnarray}
{\bf M}^{*}& = &\frac{\mbox{adj}({\bf I}-\hat{{\bf
A}}_{1})p^{**}(-1)}{d} \left({\bf C} \mu_{1} - \frac{{\bf
D}^{*}}{p^{**}(-1)}\mu_0\right)\nonumber \\ & = & \hat{{\bf
K}}_{1}p^{**}(-1)\mu_{1} - \hat{{\bf
K}}_{2}(\mu_1-\mu_2),\label{finalM*2}
\end{eqnarray}
where \[\hat{\bf K}_1=\frac{\mbox{adj}({\bf I}-\hat{\bf A}_1){\bf
    C}}{d}\;\;\; \mbox{and} \;\;\;\hat{{\bf
    K}}_{2}=\frac{\mbox{adj}({\bf I}-\hat{\bf A}_{1}){\bf
    D}^{*}}{d}.\]

\noindent Writing (\ref{finalM*2}) in scalar form 
\begin{equation}
m_1^{*} = k_{1}\mu_{1}p^{**}(-1) - k_{2}(\mu_1-\mu_2),
\end{equation}
and substituting this into (16) we obtain 
\begin{equation}
m_1^{**} = k_{1}\mu_{1}p^{*}(-1) - k_{3}(\mu_1-\mu_2).
\end{equation}

Hence in two parameter $n$-dimensional discontinuous map (\ref{normal-nd}),
a period-2 orbit will occur only when the following two
conditions are simultaneously satisfied:
\begin{eqnarray}
k_{1}p^{**}(-1)\mu_1-k_{2}(\mu_1-\mu_2)<0, \label{2fcon1}\\
k_{1}p^{*}(-1)\mu_1-k_{3}(\mu_1-\mu_2)>0.\label{2fcon2}
\end{eqnarray}
where $k_1$ and $k_2$ are respectively the first components of
$\hat{\bf K}_{1}$ and $\hat{\bf K}_{2}$, and 
\begin{equation}
k_3 = k_2
\frac{p^*(-1)}{p^{**}(-1)} + \frac{d_1}{b_1}\left(1 -
\frac{p^*(-1)}{p^{**}(-1)}\right).\label{k3}
\end{equation}

The derivation of the conditions for the existence of orbits of
periodicity greater than 2 in the $n$-dimensional discontinuous map is
very cumbersome. This is still an open problem.

Note that (\ref{2fcon1}) and (\ref{2fcon2}) give the conditions of
existence of period-2 orbit. In addition, one has to consider the
stability of the orbit in order to determine whether the period-2
orbit will actually be observable in the asymptotically stable
behavior of the system.

\section{Example: A Two Dimensional Piecewise Linear Discontinuous Map
  in Normal Form} 

The normal form for the two dimensional discontinuous map is
\begin{equation} 
\left( \begin{array}{c} x^{(n+1)} \\ y^{(n+1)} \end{array}\right)=
\left\{\begin{array}{cc} {\bf A}_1\left( \begin{array}{c} x^{(n)} \\
y^{(n)}
\end{array}\right) + {\bf C} \mu_1 , & \mbox{if}\;\; x^{(n)} < 0\\ {\bf A}_2 
\left(
\begin{array}{c} x^{(n)} \\ y^{(n)} \end{array}\right) +{\bf C} \mu_2, &
\mbox{if}\;\; x^{(n)} > 0 \end{array} \right. \label{2dmap}
\end{equation}
where \[{\bf A}_1= \left( \begin{array}{cc} \tau_L & 1 \\ -\delta_L &
0 \end{array}\right), \;\;\; {\bf A}_2= \left( \begin{array}{cc}
\tau_R & 1 \\ -\delta_R & 0 \end{array}\right), \;\;\;{\bf C} = \left(
\begin{array}{c} 1 \\ 0
\end{array}\right),\] $\tau_L$ is the trace and $\delta_L$ is the
determinant of the Jacobian matrix ${\bf A}_1$ of the system at a fixed
point in $L:=\{(x,y)\in \Re^{2}:x<0\}$ and $\tau_R$ is the trace and
$\delta_R$ is the determinant of the Jacobian matrix ${\bf A}_2$ of the
system at a fixed point in $R:=\{(x,y)\in \Re^{2}:x>0\}$. The normal
form (\ref{2dmap}) follows from that derived for piecewise smooth
continuous maps in \cite{Nusse92,cas2d}, with discontinuity added
through the parameters $\mu_1$ and $\mu_2$.

In case of the map (\ref{2dmap}), the terms $p^*(1)$,
$p^{**}(1)$, $p^*(-1)$ and
$p^{**}(-1)$ are given by
\begin{eqnarray*}
p^*(1)&=& |{\bf I}-{\bf A}_1|=\left| \begin{array}{cc} 1-\tau_L & -1 \\
  \delta_L & 1 \end{array}\right|=(1-\tau_L + \delta_L),\\ p^{**}(1)
  &=&|{\bf I}-{\bf A}_2|=\left| \begin{array}{cc} 1-\tau_R & -1 \\
  \delta_R & 1 \end{array}\right|=(1-\tau_R + \delta_R).\\ p^*(-1) &=&
  |\!-\!{\bf I}-{\bf A}_1|=\left|\! \begin{array}{cc} -1-\tau_L & -1
  \\ \delta_L & -1 \end{array}\!\right|=(1+\tau_L + \delta_L),\\
  p^{**}(-1) &=&|\!-\!{\bf I}-{\bf A}_2|=\left|\! \begin{array}{cc}
  -1-\tau_R & -1 \\ \delta_R & -1 \end{array}\!\right|=(1+\tau_R +
  \delta_R).
\end{eqnarray*}

\subsection{Inference about period-1 fixed points}

Now if we choose $p^*(1)>0$ and $p^{**}(1)>0$, i.e.,
$p^*(1)p^{**}(1)>0$, application of the results of $n$-dimensional
discontinuous map (for $n=2$) will yield the following inferences.

If we keep $\mu_2$ constant and vary $\mu_1$, then

\begin{itemize}
\item if $\mu_1 \mu_2 > 0$ then only one period-1 fixed point will exist,
\item if $\mu_1 \mu_2 < 0$ then either two period-1 fixed points will
  exist or no period-1 fixed point will exist.
\end{itemize}

We now show, using the specific 2D map (\ref{2dmap}) that the above
results are true.

\subsection{Proof of the existence of ${\bf X}_L^*$ and ${\bf X}_R^*$
  fixed points} 

If a fixed point ${\bf X}_L^*=\left( x_L^*,\; y_L^* \right)^T$ exists, it is
given by the solution of the system of equations
\begin{eqnarray}  
\left( \begin{array}{c} x_L^* \\ y_L^* \end{array}\right)& = &\left(
\begin{array}{cc} \tau_L & 1 \\ -\delta_L & 0 \end{array}\right)
\left( \begin{array}{c} x_L^* \\ y_L^* \end{array}\right) + \left(
\begin{array}{c} 1 \\ 0 \end{array}\right)\mu_1\nonumber \\
&= &\frac{\mu_1}{1-\tau_L+\delta_L}\left( \begin{array}{c} 1 \\
-\delta_L \end{array}\right).
\end{eqnarray}
Now the fixed point will exist if $x_L^*<0$, i.e., 
\begin{equation}
\frac{\mu_1}{1-\tau_L+\delta_L}= \frac{\mu_1}{p^*(1)} <0
\end{equation}
This is true if
\begin{eqnarray}
\{\tau_L \!>\! 1+\delta_L \;\;\mbox{or equivalently}\;\;
p^*(1)\!<\!0\}\; & \;\;
\mbox{for} & \;\;\mu_1 \!>\! 0, \label{2d1}\\ 
\{\tau_L \!<\! 1+\delta_L  \;\;\mbox{or equivalently}\;\;
p^*(1)\!>\!0\} \; & \;\;
\mbox{for} &\;\; \mu_1 \!<\! 0.\label{2d2}
\end{eqnarray}

\begin{figure}[tbh]
\centering
\includegraphics[width=0.7\columnwidth]{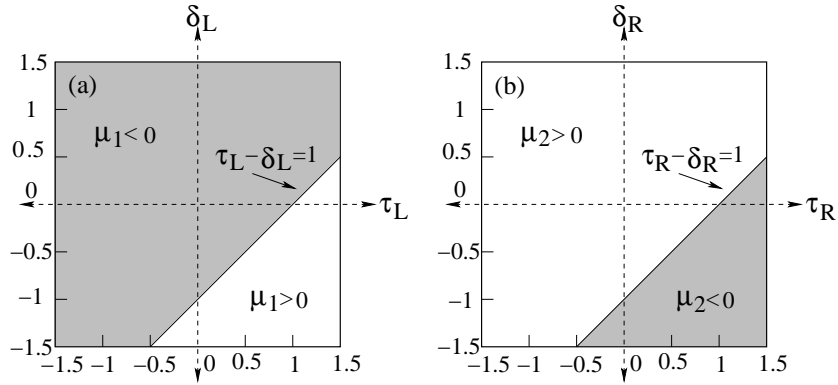}
\caption{\label{fig2label}(a) Regions of existence of the ${\bf
X}_L^*$ fixed point as $\mu_1$ is varied through 0. The fixed point
exists in the shaded region if $\mu_1<0$, and in the unshaded region
if $\mu_1>0$. (b) Regions of existence of the ${\bf X}_R^*$ fixed
point as $\mu_2$ is varied through 0. The fixed point exists in the
shaded region if $\mu_2<0$, and in the unshaded region if
$\mu_2>0$. }
\end{figure}

The above conditions (\ref{2d1}) and (\ref{2d2}) are depicted
graphically in Fig.~\ref{fig2label}(a).

In a similar manner it can be shown that the fixed point ${\bf X}_R^*$
will exist if
\begin{equation}
\frac{\mu_2}{1-\tau_R+\delta_R}= \frac{\mu_2}{p^{**}(1)} >0
\end{equation}
This is true if
\begin{eqnarray}
\{\tau_R \!>\! 1+ \delta_R \;\; \mbox{or equivalently}\;\; p^{**}(1)\!<\!0\}
& \;\;\mbox{for} \;\;\; \mu_2 \!<\! 0, \label{2d3}\\
\{\tau_R \!<\! 1+ \delta_R \;\; \mbox{or equivalently}\;\; p^{**}(1)\!>\!0\} &
\;\;\mbox{for} \;\;\; \mu_2 \!>\! 0.\label{2d4}
\end{eqnarray}

The above conditions (\ref{2d3}) and (\ref{2d4}) are depicted
graphically in Fig.~\ref{fig2label}(b).

\begin{figure}[tbh]
\centering
\includegraphics[width=0.5\columnwidth]{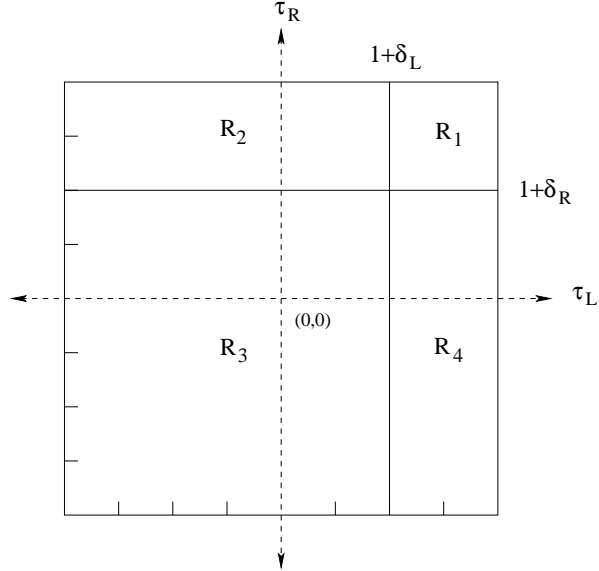}
\caption{Partitioning of the parameter space where different
  bifurcations should be observed.}
\label{fig4label}
\end{figure}

The conditions (\ref{2d1}), (\ref{2d2}), (\ref{2d3}), and (\ref{2d4})
divide the $(\tau_L,\tau_R)$ parameter space into four regions $R_1,
R_2, R_3$ and $R_4$ as shown in
Fig.~\ref{fig4label}. These regions correspond to the four rows in
Table~\ref{table1}.


In the region $R_1$, since $\tau_L > 1+\delta_L$
and $\tau_R > 1+\delta_R$, the period one fixed points ${\bf X}_L^*$
and ${\bf X}_R^*$ are not stable. Therefore, even though they exist,
the trajectory diverges to infinity.

 
\begin{figure}[tbh]
\centering
\includegraphics[width=0.5\columnwidth]{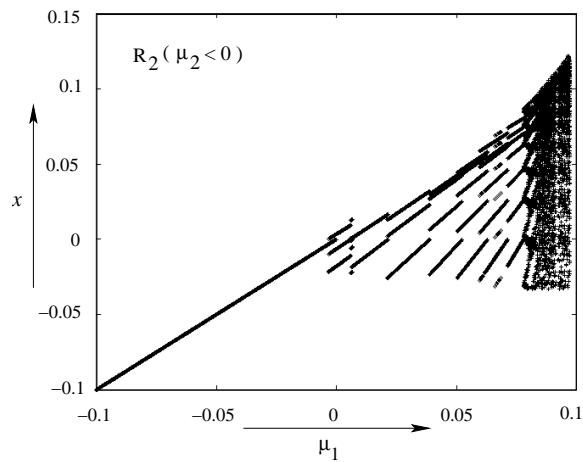}
\caption{A one-parameter bifurcation diagram with $\mu_1$ as the
  bifurcation parameter and $\mu_2=-0.025$. The other parameters
  $\tau_L=0.3,\; \delta_L=0.3,\; \tau_R=1.5$ and $\delta_R=0.3$ are in ${\bf
  R}_2$.}
\label{2regmune}
\end{figure}


\begin{figure}[tbh]
\centering
\includegraphics[width=0.5\columnwidth]{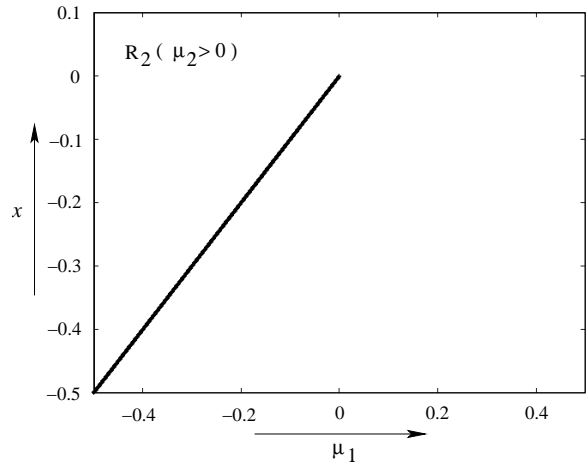}
\caption{A one-parameter bifurcation diagram with $\mu_1$ as the
  bifurcation parameter and $\mu_2=0.025$ . The other parameters
  $\tau_L=0.3,\; \delta_L=0.3,\; \tau_R=1.5$ and $\delta_R=0.3$ are in ${\bf
  R}_2$.}
\label{2regmu2pe}
\end{figure}

In $R_2$, suppose we keep $\mu_2$ fixed at a negative value, and vary
$\mu_1$. Table~\ref{table1} shows that the period-1 fixed point ${\bf
X}_L^*$ will exist so long as $\mu_1<0$, and the fixed point ${\bf
X}_R^*$ will exist for the whole range of $\mu_1$. Now, if we choose
the parameters such that $-(1+\delta_L)<\tau_L < 1+\delta_L$ and
$\tau_R > 1+\delta_R$, then ${\bf X}_L^*$ is stable and ${\bf X}_R^*$
is unstable. The representative bifurcation diagram for such a
situation is shown in Fig.~\ref{2regmune}, which shows the period-1
orbit for $\mu_1<0$ which disappears for $\mu_2>0$. If
we take $\mu_2>0$ and vary $\mu_1$, then we infer from
Table~\ref{table1} that ${\bf X}_L^*$ will exist (and is stable) for
$\mu_1<0$, but no fixed point will exist for $\mu_1>0$. This is what
we see in the bifurcation diagram of Fig.~\ref{2regmu2pe}.


\begin{figure}[tbh]
\centering
\includegraphics[width=0.5\columnwidth]{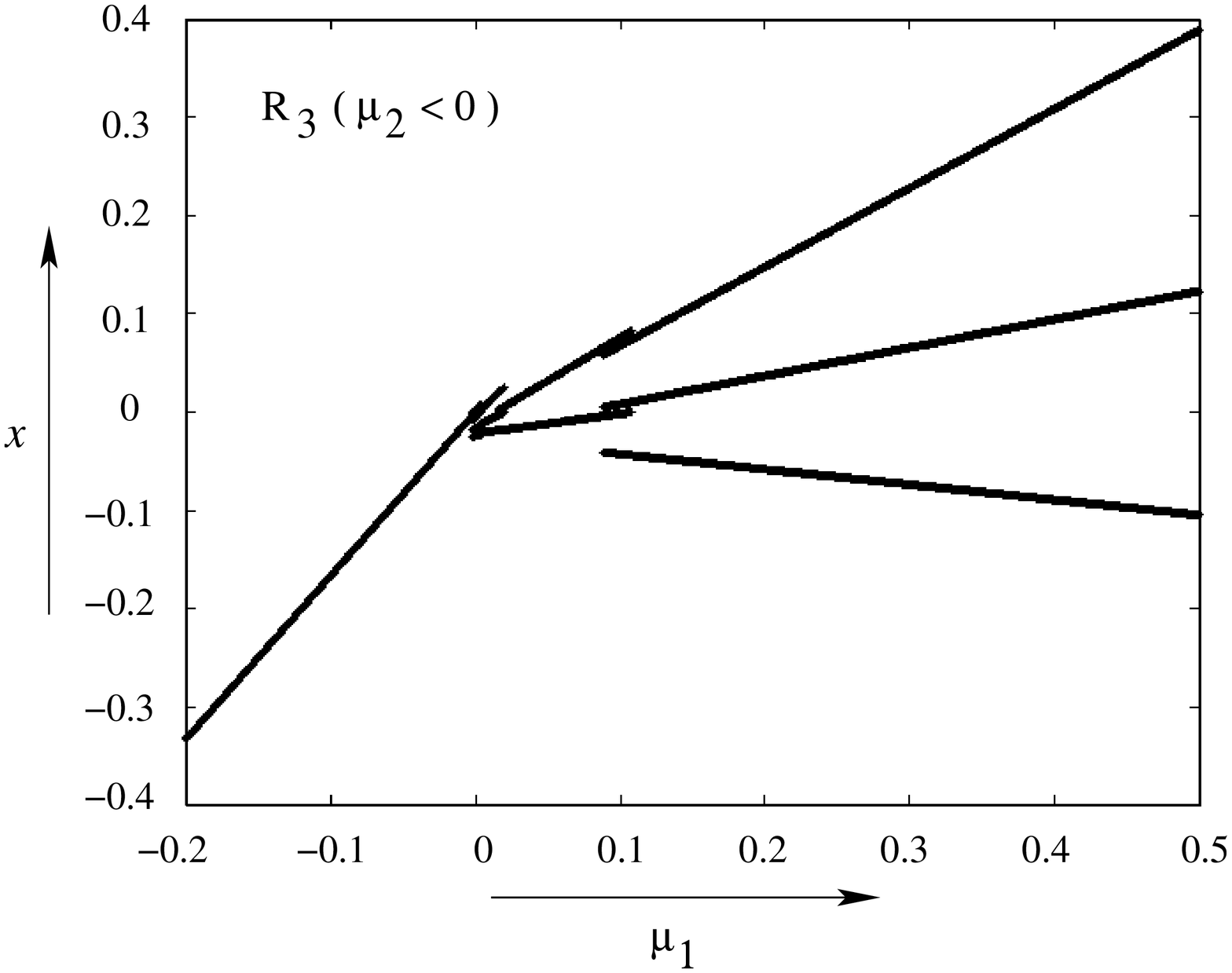}
\caption{A one-parameter bifurcation diagram with $\mu_1$ as the
  bifurcation parameter and $\mu_2=-0.025$ . The other parameter
  $\tau_L=0.7,\; \delta_L=0.3,\; \tau_R=0.3$ and $\delta_R=0.3$ are in
  ${\bf R}_3$.}
\label{3regmune}
\end{figure}


\begin{figure}[tbh]
\centering
\includegraphics[width=0.5\columnwidth]{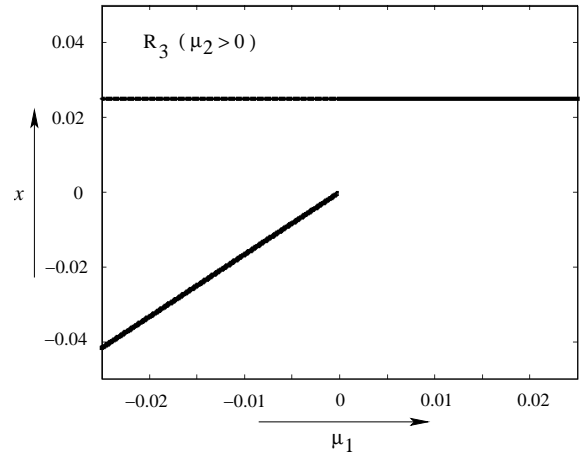}
\caption{A one-parameter bifurcation diagram with $\mu_1$ as the
  bifurcation parameter and $\mu_2=0.025$ . The other
  parameters $\tau_L=0.7,\; \delta_L=0.3,\; \tau_R=0.3$ and
  $\delta_R=0.3$ are in ${\bf R}_3$.}
\label{3regmu2pe}
\end{figure}

In the region $R_3$, suppose we fix $\mu_2<0$ and vary
$\mu_1$. Table~\ref{table1} shows that for $\mu_1<0$, ${\bf X}_L^*$
will exist but ${\bf X}_R^*$ will not. For $\mu_1>0$ none of the fixed
points will exist. Now, if the parameters are chosen such that
$-(1+\delta_L)<\tau_L < 1+\delta_L$, then a stable period-1 behavior
will be observed so long as $\mu_1<0$. This is shown in
Fig.~\ref{3regmune}. On the other hand, if $\mu_2$ is chosen to be
positive, for $\mu_1<0$ both the fixed points will exist, and for
$\mu_1>0$ ${\bf X}_R^*$ will exist but ${\bf X}_L^*$ will not. If the
parameters are such that $-(1+\delta_R)<\tau_R < 1+\delta_R$, then
${\bf X}_R^*$ will occur for the whole parameter range, while ${\bf
X}_L^*$ will occur only for $\mu_1<0$. This is shown in the simulated
bifurcation diagram of Fig.~\ref{3regmu2pe}.


\begin{figure}[tbh]
\centering
\includegraphics[width=0.5\columnwidth]{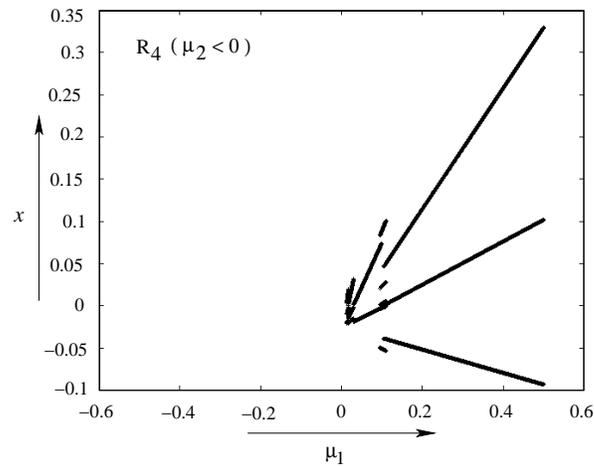}
\caption{A one-parameter bifurcation diagram with $\mu_1$ as the
  bifurcation parameter and $\mu_2=-0.025$ . The other parameters
  $\tau_L=1.5,\; \delta_L=0.3,\; \tau_R=0.3$ and $\delta_R=0.3$ are in ${\bf
  R}_4$.}
\label{4regmune}
\end{figure}


\begin{figure}[tbh]
\centering
\includegraphics[width=0.5\columnwidth]{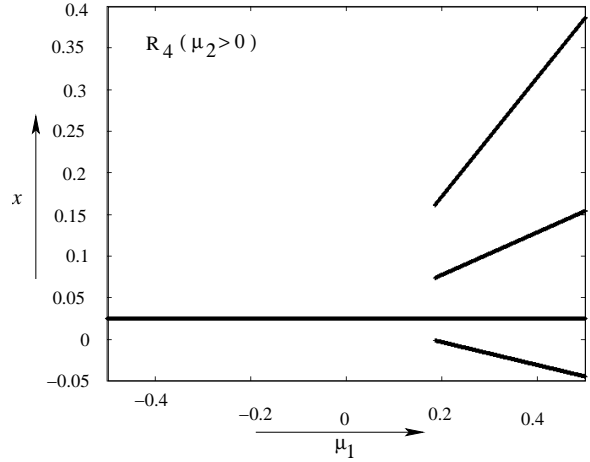}
\caption{A one-parameter bifurcation diagram with $\mu_1$ as the
  bifurcation parameter and $\mu_2=0.025$ . The other parameters
  $\tau_L=1.5,\; \delta_L=0.3,\; \tau_R=0.3$ and $\delta_R=0.3$ are in
  ${\bf R}_4$.}
\label{4regmu2pe}
\end{figure}

Similarly in $R_4$, suppose we keep $\mu_2$ fixed at a negative value,
and vary $\mu_1$. Table~\ref{table1} shows that the none of the fixed
points will exist so long as $\mu_1<0$, and for $\mu_1>0$ the fixed
point ${\bf X}_L^*$ will exist (and will be unstable) but ${\bf
X}_R^*$ will not exist. The representative bifurcation diagram for
such a situation is shown in Fig.~\ref{4regmune}. When $\mu_2<0$, the
period-1 fixed point is not observed over the whole parameter range,
but the existence of the unstable fixed point ${\bf X}_L^*$ causes
high periodic orbits to occur for $\mu_1>0$. On the other hand, if
$\mu_2$ is chosen to be positive, for $\mu_1<0$ the fixed point ${\bf
X}_R^*$ will exist but ${\bf X}_L^*$ will not exist and for $\mu_1>0$
both the fixed points will exist. If the parameters are chosen such
that $-(1+\delta_R)<\tau_R < 1+\delta_R$, then ${\bf X}_R^*$ is stable
and will occur for the whole parameter range and ${\bf X}_L^*$ is
unstable. This is shown in the simulated bifurcation diagram of
Fig.~\ref{4regmu2pe}.

\subsection{Conditions for existence of period-2 orbits}

The conditions for existence of period-2 orbits can be obtained from
(\ref{2fcon1}) and (\ref{2fcon2}). First we have to find the values of
$k_1$, $k_2$ and $k_3$.

From Section~\ref{secIII}, we have \[\hat{{\bf A}}_{1} = {\bf A}_{1} -
\frac{1}{p^{**}(-1)} \left(
\begin{array}{llcl} b_1^{*} & 0 \\ b_2^{*} & 0 \end{array}\right) = \left(
\begin{array}{llcl} \tau_L - \frac{b_1^{*}}{p^{**}(-1)} & 1 \\ -\delta_L - 
\frac{b_2^{*}}{p^{**}(-1)} & 0 \end{array}\right),\]

\begin{eqnarray*}
{\bf B}^*&=&\left(
\begin{array}{c} b_1^* \\ b_2^* \end{array}\right) = 
(-1)^{2}(\mbox{adj}({\bf I}+{\bf A}_2))[{\bf A}_1-{\bf A}_2]_1 \\ & = &
\mbox{adj}\left(
\begin{array}{llcl} 1+\tau_R & 1 \\ -\delta_R & 1 \end{array}\right)
\left(\begin{array}{c} \tau_L -\tau_R \\ -\delta_L +\delta_R
\end{array}\right)\\ & = & \left(\begin{array}{c} \tau_L -\tau_R +\delta_L -
\delta_R\\ \delta_R (1+\tau_L)-\delta_L (1+\tau_R)\end{array}\right),
 \end{eqnarray*}

\[d=|{\bf I}-\hat{{\bf A}}_{1}|p^{**}(-1)=(1+\delta_L)(1+\delta_R)-\tau_L 
\tau_R\]
and
\begin{eqnarray*}
{\bf D}^{*}&=&(-1)^{2}(\mbox{adj}({\bf I}+{\bf A}_2)){\bf C} \\ &=& \left(
\begin{array}{llcl} 1 & -1 \\ \delta_R & 1+\tau_R \end{array}\right) \left(
\begin{array}{c} 1 \\ 0 \end{array}\right)= \left(
\begin{array}{c} 1 \\ \delta_R \end{array}\right).
\end{eqnarray*}

Now with the help of above values, we find
\begin{eqnarray*}
\hat{{\bf K}}_{1}&=&\frac{\mbox{adj}({\bf I}-\hat{{\bf A}}_{1}){\bf C}}{d} =
\frac{1}{d}\left(
\begin{array}{c} 1 \\ -\delta_L-\frac{b^*_2}{p^{**}(-1)} \end{array}\right),\\
\hat{{\bf K}}_{2} &=& \frac{\mbox{adj}({\bf I}-\hat{{\bf A}}_{1}){\bf
D}^*}{d} = \frac{1}{d}\left(
\begin{array}{c} 1+\delta_R \\ -\delta_L-\frac{b^*_2}{p^{**}(-1)}+
\delta_R-\delta_R\tau_L+\delta_R\frac{b^*_1}{p^{**}(-1)}
\end{array}\right).
\end{eqnarray*}

\noindent Substituting the parameters, we obtain

\begin{eqnarray*}
k_1 = \frac{1}{(1+\delta_L)(1+\delta_R)-\tau_L \tau_R}\\
k_2 = \frac{1+\delta_R}{(1+\delta_L)(1+\delta_R)-\tau_L \tau_R}
\end{eqnarray*}
and from (\ref{k3})
\[
k_3 = \frac{\tau_L}{(1+\delta_L)(1+\delta_R)-\tau_L \tau_R}.
\]

Using these expressions, (\ref{2fcon1}) and (\ref{2fcon2})
give
\begin{equation}
\frac{\tau_R \mu_1+(1+\delta_R)\mu_2}{(1+\delta_L)(1+\delta_R)
-\tau_L \tau_R} < 0 \label{p2cond-a}
\end{equation}
\begin{equation}
\frac{(1+\delta_L)\mu_1+\tau_L\mu_2}{(1+\delta_L)(1+\delta_R)
-\tau_L \tau_R} > 0\label{p2cond-b}
\end{equation}
as the conditions of existence of the period-2 orbit.

We now show that these are indeed the conditions of the existence of
the period-2 fixed points in the 2D map.

\subsection{Proof of the conditions of existence of period-2 orbits
  using (\ref{2dmap})}

Since the system is linear in each side of the border, period two (or
a higher period) fixed points cannot exist with all points in
$L$ or all points in $R$. However in some regions of the
parameter space, a period two fixed point may exist with one point in
$L$ and one point in $R$. The conditions for
existence of period-2 orbit are obtained by solving
\begin{equation}  
\left( \begin{array}{c} m_1^{**} \\ m_2^{**} \end{array}\right) = \left(
\begin{array}{cc} \tau_L & 1 \\ -\delta_L & 0 \end{array}\right)
\left( \begin{array}{c} m_1^* \\ m_2^* \end{array}\right) + \left(
\begin{array}{c} 1 \\ 0 \end{array}\right)\mu_1
\end{equation}
\begin{equation}  
\left( \begin{array}{c} m_1^* \\  m_2^* \end{array}\right) = \left(
\begin{array}{cc} \tau_R & 1 \\ -\delta_R & 0 \end{array}\right)
\left( \begin{array}{c} m_1^{**} \\ m_2^{**} \end{array}\right) + \left(
\begin{array}{c} 1 \\ 0 \end{array}\right)\mu_2
\end{equation}
where $ m_1^* < 0$ and $ m_1^{**} > 0$.

Now solving the above two equations we find
\begin{equation}
 m_1^* = \frac{\tau_R \mu_1+(1+\delta_R)\mu_2}{(1+\delta_L)(1+\delta_R)
-\tau_L \tau_R} < 0
\end{equation}
\begin{equation}
m_1^{**} = \frac{(1+\delta_L)\mu_1+\tau_L\mu_2}{(1+\delta_L)(1+\delta_R)
-\tau_L \tau_R} > 0
\end{equation}
These are the same as those obtained from the theory for
$n$-dimensional systems.

This implies that if 
\[\tau_L \tau_R > (1+\delta_L)(1+\delta_R)\] then the period-2 orbit
will exist when
\begin{eqnarray} 
-\frac{\tau_R}{(1+\delta_R)} < \frac{\mu_2}{\mu_1} <
-\frac{(1+\delta_L)}{\tau_L} 
\end{eqnarray}
and if
\[\tau_L \tau_R < (1+\delta_L)(1+\delta_R) \]
then the period-2 orbit will exist when
\begin{equation} 
-\frac{\tau_R}{(1+\delta_R)} > \frac{\mu_2}{\mu_1} >
-\frac{(1+\delta_L)}{\tau_L} .
\end{equation}

\begin{figure}[tbh]
\centering
\includegraphics[width=0.7\columnwidth]{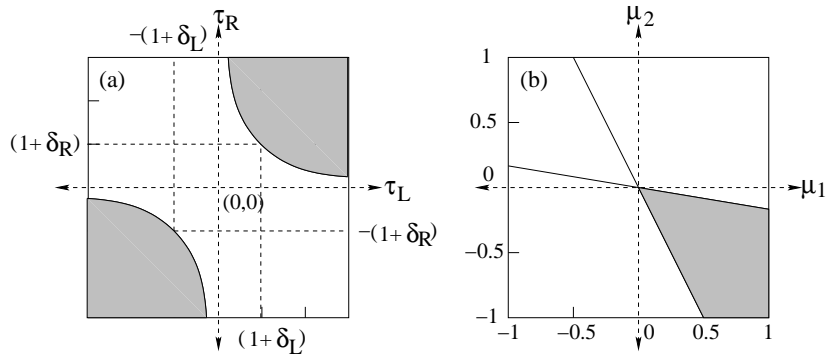}
\caption{(a) Regions of existence of period-2 points depending on
$\tau_L$ and $\tau_R$.  The period-2 points exist in the shaded
region if $\tau_L \tau_R > (1+\delta_L)(1+\delta_R)$ and in the
unshaded region if $\tau_L \tau_R < (1+\delta_L)(1+\delta_R)$.  (b)
Regions of existence of period-2 point in $\mu_1 \!-\!\mu_2$ parameter
space. The period-2 points exist in the shaded regions for a set of
particularly chosen $\tau_L,\; \tau_R,\; \delta_L\; and\;
\delta_R$.}\label{fig1label3}
\end{figure}

These two conditions are shown in Fig.~\ref{fig1label3}(a) as the shaded
and unshaded regions respectively in the $\tau_L\!-\!\tau_R$ space.
Now, for particular values of the parameters $\tau_L=0.7$, $\delta_L=0.3$,
$\tau_R=0.2$ and $\delta_R=0.3$  the region of existence of period-2 are  
depicted in  the $\mu_1\!-\!\mu_2$ space in Fig.~\ref{fig1label3}(b).


\begin{figure}[tbh]
\centering
\includegraphics[width=0.5\columnwidth]{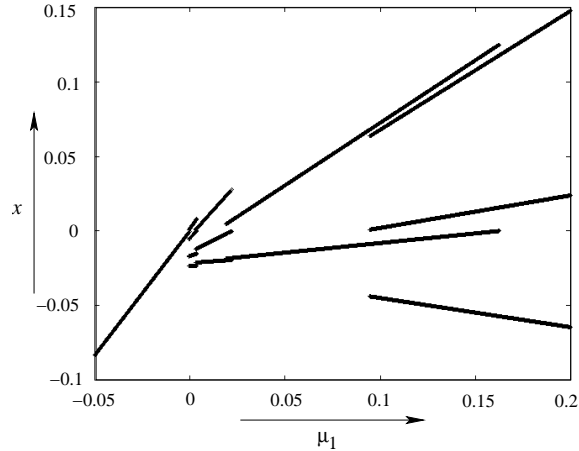}
\caption{A representative one-parameter bifurcation diagram with
  $\mu_1$ as the bifurcation parameter. The other parameters are:
$\tau_L=0.7$, $\delta_L=0.3$, $\tau_R=0.2$, $\delta_R=0.3$, $\mu_2=-0.025$.}
\label{fig1label4}
\end{figure}

These equations allow one to calculate the parameter range of
existence of the period-2 orbit, when one of the parameters is
varied. To illustrate this, we show the bifurcation diagram in
Fig.~\ref{fig1label4}, where $\mu_1$ varied and all the other
parameters are kept fixed. The stability of the period-2 fixed point
is given by the eigenvalues of the matrix ${\bf A}_1{\bf A}_2$.  For
the chosen parameter values, these are $\lambda_1=-0.23+0.19i$ and
$\lambda_2=-0.23-0.19i $. This implies that if the period-2 fixed
point exists, it will be stable. Now from (\ref{p2cond-a}) and
(\ref{p2cond-b}) we find that the period-2 orbit should exist between
$\mu_1=0.01346$ and $\mu_1=0.1625$. Again, since $\mu_2<0$,
Table~\ref{table1} predicts that the period-1 orbit should exist so
long as $\mu_1<0$. The bifurcation diagram clearly shows that the
above predictions are true.


\section{Conclusions}

In view of the fact that a large number of systems of practical
interest give rise to discontinuous maps, it has become necessary to
develop a comprehensive theory of bifurcations in such systems. In this
paper we have taken the first step in that direction by extending
Feigin's method that was earlier developed in the context of continuous
maps. We have obtained the conditions of existence of the period-1 and
period-2 fixed points in general $n$-dimensional discontinuous maps,
and have demonstrated the application of the derived results in a
piecewise linear two dimensional discontinuous map.

It may be noted that the results obtained in this paper speak only of
the existence of period-1 and period-2 fixed points. In order to
obtain information about the occurrence of high-periodic orbits and
chaos, one has to consider specific dimension of the system, which
cannot be obtained in the general $n$-dimensional context.

\ack

One of the author (P.S.Dutta) thanks CSIR India for financial support
in the form of a CSIR-JRF. This work was supported in part by the
BRNS, Department of Atomic Energy, Government of India under project
no. 2003/37/11/BRNS.


\end{document}